# 量子金鑰分配實作挑戰

Abel C. H. Chen
*Information & Communications Security Laboratory,*
*Chunghwa Telecom Laboratories*
Taoyuan, Taiwan
Email: chchen.scholar@gmail.com; ORCID: 0000-0003-3628-3033

*摘要*—近年來量子計算技術逐漸成熟，並且開始被實際應用到各個領域。其中一個重要領域為網路通訊加密，通過量子金鑰分配(Quantum Key Distribution, QKD)協商出通訊雙方共享秘密(Shared Secret)，再運用此共享秘密產製對稱式金鑰，後續可以應用於加解密需求。本研究主要聚焦在真實量子電腦環境實作和比較兩個知名的量子金鑰分配協定(包含 BB84 協定和 E91 協定)，並且提出基於 SX Gate 操作產生均勻量子疊加態，通過量子疊加態、量子糾纏態等量子物理特性讓通訊雙方可以取得共享秘密，同時避免被攻擊者竊取共享秘密。本研究採用 IBM Quantum Platform 實作，通過在真實量子電腦上實證 BB84 協定和 E91 協定的可行性，分別從熵(entropy)驗證、獨立且同分布(Independent and Identically Distributed, IID)驗證、以及錯誤率驗證等指標來評估。

*關鍵字*—量子金鑰分配、BB84 協定、E91 協定、量子疊加態、量子糾纏態

## I. 前言

隨著量子計算技術的普及，許多領域開始著手把量子計算技術應用在各行各業[1],[2]，並且也開始制定相關標準[3],[4]。其中，網路通訊加密也開始考慮結合量子金鑰分配(Quantum Key Distribution, QKD)，來作為另一種量子安全(Quantum Safe)的解決方案[5],[6]。因此，近幾年陸續有研究開始討論在傳輸層安全性協定(Transport Layer Security, TLS)結合量子金鑰分配[7],[8]。有鑑於此，未來部署量子安全網路通訊時，量子金鑰分配將會是重要的技術之一。

因此，本研究將聚焦在目前主流的量子金鑰分配協定(包含 BB84 協定[9]和 E91 協定[10])，並且分別在真實量子電腦環境上實現，驗證其量子物理特性，運用量子疊加態、量子糾纏態來保護通訊雙方共享秘密(Shared Secret)。其中，將深入探索在真實量子電腦環境進行量測的誤差，並且驗證分享共享秘密的隨機性，從而驗證 BB84 協定和 E91 協定的可行性。本研究貢獻條列如下：

- 本研究提出通過操作 SX Gate 操作產生均勻量子疊加態，為實現 BB84 協定和 E91 協定提供另一種解決方案。

- 本研究在 IBM Quantum Platform 實作 BB84 協定和 E91 協定，運用真實量子電腦實證不同的量子金鑰分配協定的效能，以及量子物理特性的影響。

- 本研究通過實現真實量子電腦環境下的 BB84 協定和 E91 協定，分別產生共享秘密，並且討論隨機位元熵(entropy)驗證和獨立且同分布(Independent and Identically Distributed, IID)驗證結果。

- 本研究實證 BB84 協定和 E91 協定在協商過程中和量測過程中可能產生的誤差，並且驗證錯誤率，用以說明量子金鑰分配協定的可行性。

本文主要分為 6 個小節。其中，第 II 節介紹量子疊加態和量子糾纏態。第 III 節和第 IV 節分別介紹 BB84 協定和 E91 協定的概念和量測結果。第 V 節提供真實量子電腦環境上的實測結果，並深入討論觀察到的現象。最後，第 VI 節總結本研究發現和貢獻，以及討論未來可行的研究方向。

## II. 量子態

經典位元(Classic Bit)與量子位元(Quantum Bit, Qubit)有著顯著差異；其中，經典位元只能是 0 或 1，但量子位元存在量子疊加態，在量測前有機率是 0，也有機率是 1。為了以視覺化方式呈現量子疊加態，本研究採用 IBM Quantum Platform 的 Composer 功能，展示量子電路及其對應的量測結果機率。圖 1 顯示只有 1 個量子位元(q[0])的情況下，並且預設值為|0⟩，所以不做任何操作下直接量測(即 ![icon])將得到|0⟩的機率是 100%。本研究將於第 II.A 節介紹量子疊加態，以及第 II.B 節介紹量子糾纏態。

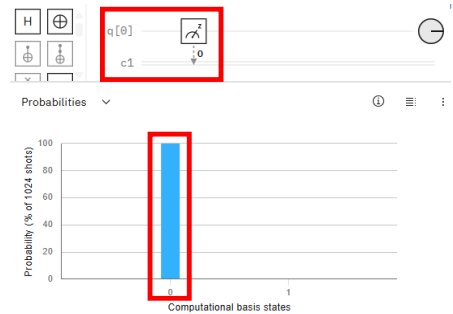

Fig. 1. The default quantum state.

### A. 量子疊加態

在密碼學領域，為提升安全性，目標將會是隨機得到 0 和 1 的機率為各 50%，可以達到較高的熵值。因此，為達到此效果，則主要在量子電腦建立均勻量子疊加態。其中，建立均勻量子疊加態的方式有很多種[11]，本研究提供採用 Hadamard Gate 操作(即 ![H])的方式來建立均勻量子疊加態，如圖 2 所示。可以觀察到，預設值為 100%是|0⟩ (如圖 1 所示)，但在操作 Hadamard Gate 後，轉換為 50%是|0⟩、50%是|1⟩。因此，在此操作後，再量測得到位元值後，可以得到真隨機位元值，並且具有較高的熵值。除此之外，本研究觀察通過操作 SX Gate (即 ![icon])或 Inversed SX Gate (即 ![icon])也可以建立均勻量子疊加態，如圖 3 所示。因此，後續在 BB84 協定和 E91 協定的控制訊號(或稱"測量基")，本研究也將採用 SX Gate 和 Inversed SX Gate 的組合來產生均勻量子疊加態。

### B. 量子糾纏態

量子糾纏態的量子物理特性，讓兩個處於量子糾纏態的量子位元，當其中一個量子位元進行量測後，也將會

對另一個量子位元的量測結果造成影響[12]。並且在量測前，持有量子位元的雙方都無法預測量測結果的值會是多少。本節以貝爾態(Bell state)(一種量子糾纏態)為例進行說明，在案例中假設有兩個量子位元，分別是 q[0]和 q[1]。首先，對 q[0]操作 Hadamard Gate，再對 q[0]和 q[1]操作 CNOT Gate (即⊕)，設定 q[0]為控制位元，而 q[1]為目標位元，如圖 4 所示。可以觀察到，預設值為 100%是|00⟩ (如圖 1 所示)，但在操作 Hadamard Gate 和 CNOT Gate 後，|q[1]q[0]⟩轉換為 50%是|00⟩、50%是|11⟩。因此，在此操作後進行量測，當 q[0]量測得到|0⟩，則 q[1]也一定量測得到|0⟩；當 q[0]量測得到|1⟩，則 q[1]也一定量測得到|1⟩。

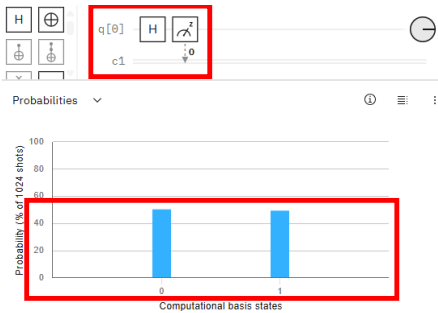

Fig. 2. The Hadamard gate-based uniform quantum superposition state.

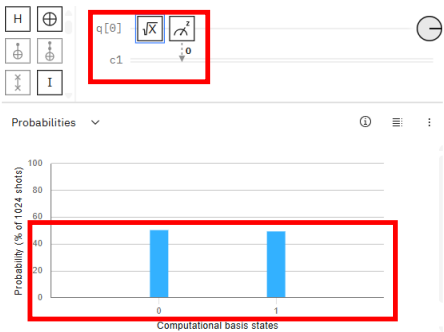

Fig. 3. The SX gate-based uniform quantum superposition state.

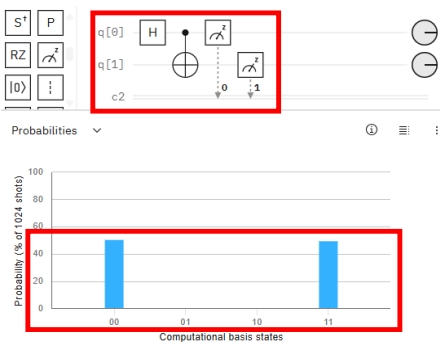

Fig. 4. The Bell state (i.e. one of quantum entanglement stage).

### III. BB84 協定

第 III.A 節介紹 BB84 協定的流程及步驟，第 III.B 節介紹本研究採用的控制訊號(或稱"測量基")。然後，第 III.C 節和第 III.D 節分別介紹通訊雙方選擇相同或不同控制訊號的結果。最後，第 III.E 節提供案例和討論。

#### A. 步驟

本節假設 BB84 協定的傳輸雙方分別是 Alice 和 Bob，並且由 Alice 擔任發起端。首先，Alice 將隨機設定每個量子位元的初始值，由於每個量子位元預設值為|0⟩，所以 Alice 將對部分量子位元操作 NOT Gate (即⊕)，讓該些量子位元初始值為|1⟩ (如圖 5❶所示)。然後，Alice 隨機選擇她的控制訊號，並且對每個量子位元進行操作，部分量子位元將會處於均勻量子疊加態(如圖 5❷所示)，之後再通過量子通道傳送量子位元給 Bob (如圖 5❸所示)。當 Bob 收到量子位元後，隨機選擇他的控制訊號，並且對每個量子位元進行操作(如圖 5❹所示)，之後再量測每個量子位元(如圖 5❺所示)。最後，再由 Bob 通過經典通道傳送他的控制訊號序列給 Alice (如圖 5❻所示)，之後 Alice 比對雙方的控制訊號序列，並且把比對結果通過經典通道傳送給 Bob (如圖 5❼所示)。後續雙方可以取得具有相同控制訊號的量子位元值作為協商後的共享秘密(如圖 5❽所示)。詳細數學證明可以參考[13]。

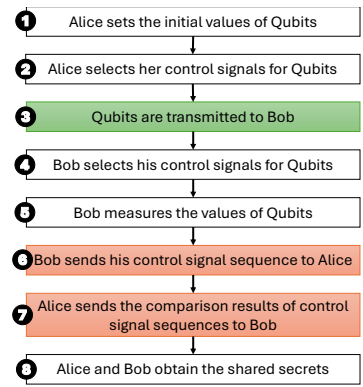

Fig. 5. The steps of the BB84 protocol.

#### B. 控制訊號(或稱"測量基")

本節採用兩種設計方法：(1).基於 Hadamard Gate 和(2). 基於 SX Gate (本研究提出方法)。

1) 基於 Hadamard Gate：採用 Hadamard Gate 和不採用 Hadamard Gate 兩種可選的控制訊號。Alice 和 Bob 可以對每個量子位元隨機選擇是否採用 Hadamard Gate 和不採用 Hadamard Gate。以圖 6 為例，首先，Alice 對量子位元操作 NOT Gate (即⊕)，讓量子位元初始值為|1⟩，之後選擇對量子位元採用 Hadamard Gate 操作(即 H)，所以量子位元將會是均勻量子疊加態。之後，Bob 也選擇對量子位元採用 Hadamard Gate 操作(即 H)，所以量子位元還原為|1⟩。因此，之後量測時會是 100%是|1⟩，再通過控制訊號序列比對結果可知，雙方對該量子位元採用相同的控制訊號，所以該量子位元的量測結果可用。

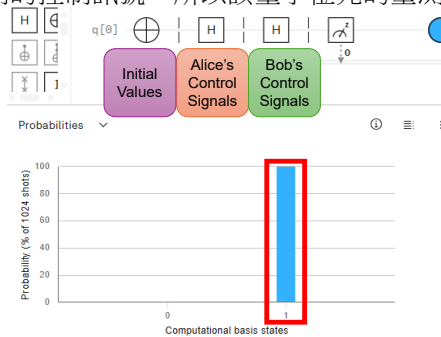

Fig. 6. The BB84 protocol based on Hadamard gates.

*2) 基於 SX Gate*：對 Alice 有採用 SX Gate 和不採用 SX Gate 兩種可選的控制訊號；對 Bob 有採用 Inversed SX Gate 和不採用 Inversed SX Gate 兩種可選的控制訊號。Alice 和 Bob 可以隨機選擇各自的控制訊號對每個量子位元進行操作。以圖 7 為例，首先，Alice 對量子位元操作 NOT Gate (即 ⊕)，讓量子位元初始值為|1⟩，之後選擇對量子位元採用 SX Gate 操作(即 √X)，所以量子位元將會是均勻量子疊加態。之後，Bob 也選擇對量子位元採用 Inversed SX Gate 操作(即 √X†)，所以量子位元還原為|1⟩。因此，之後量測時會是 100%是|1⟩，再通過控制訊號序列比對結果可知，雙方對該量子位元採用相同的控制訊號，所以該量子位元的量測結果可用。

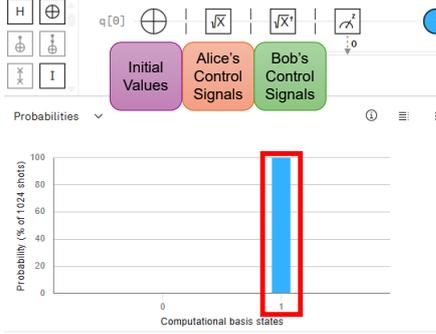

Fig. 7. The BB84 protocol based on SX gates and inversed SX gates.

### C. 通訊雙方選擇相同控制訊號

本節介紹當 Alice 和 Bob 通訊雙方選擇相同控制訊號後的量測結果，如圖 8 所示。其中，假設 Alice 分別對量子位元 q[0] 和 q[2] 操作 NOT Gate，讓量子位元|q[3]q[2]q[1]q[0]⟩初始值為|0101⟩。Alice 分別對量子位元 q[3]和 q[2]選擇不採用 Hadamard Gate，並且分別對量子位元 q[1]和 q[0]選擇採用 Hadamard Gate，然後傳送量子位元|01UU⟩，U 表示均勻量子疊加態。之後，Bob 分別對量子位元 q[3]和 q[2]選擇不採用 Hadamard Gate，並且分別對量子位元 q[1]和 q[0]選擇採用 Hadamard Gate，所以 Bob 量測後可以得到|0101⟩。由於 Alice 和 Bob 在每個量子位元都採用相同的控制訊號，所以 Alice 和 Bob 在每個量子位元都可以得到一致的量測結果。

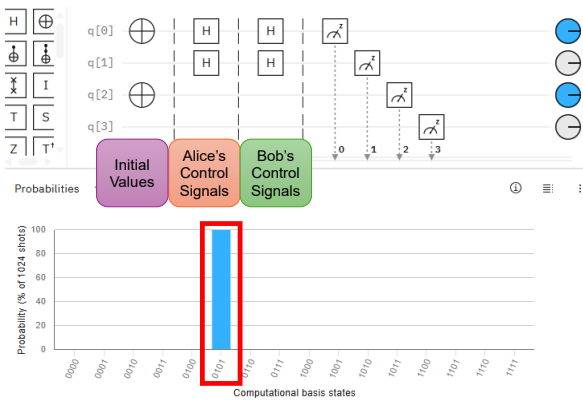

Fig. 8. Selection of the same control signals by both communication parties in the BB84 protocol.

### D. 通訊雙方選擇不同控制訊號

本節介紹當 Alice 和 Bob 通訊雙方選擇不同控制訊號後的量測結果，如圖 9 所示。其中，假設 Alice 分別對量子位元 q[0] 和 q[2] 操作 NOT Gate，讓量子位元|q[3]q[2]q[1]q[0]⟩初始值為|0101⟩。Alice 分別對量子位元 q[2]和 q[1]選擇不採用 Hadamard Gate，並且分別對量子位元 q[3]和 q[0]選擇採用 Hadamard Gate，然後傳送量子位元|U10U⟩，U 表示均勻量子疊加態。之後，Bob 分別對量子位元 q[3]和 q[0]選擇不採用 Hadamard Gate，並且分別對量子位元 q[2]和 q[1]選擇採用 Hadamard Gate，所以 Bob 量測後可以得到|UUUU⟩。由於 Alice 和 Bob 在每個量子位元都採用不同的控制訊號，所以 Alice 和 Bob 在每個量子位元都無法確定能得到一致的量測結果。

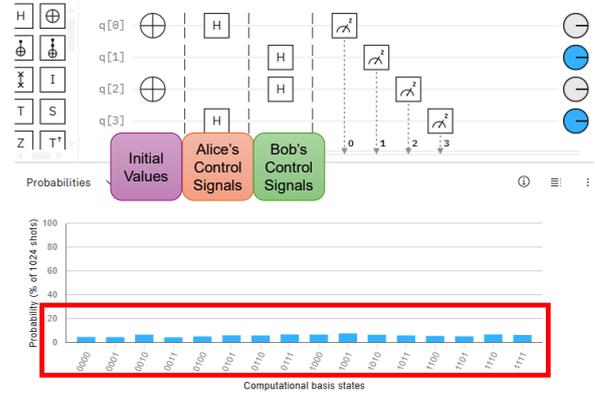

Fig. 9. Selection of the different control signals by both communication parties in the BB84 protocol.

### E. 案例與討論

在實際部署 BB84 協定時，Alice 隨機對平均一半的量子位元操作 NOT Gate，讓量子位元初始值有 50%是|0⟩，有 50%是|1⟩，讓熵值(亂度)較大。此外，控制訊號的部分，如果都選擇操作 Hadamard Gate 雖然可以讓全部量子位元在傳輸過程中都是均勻量子疊加態，但攻擊者只需對全部量子位元都 Hadamard Gate 即可還原。因此，為了提升傳輸安全和增加傳輸值的亂度，Alice 隨機對平均一半的量子位元操作 Hadamard Gate，另一半不操作 Hadamard Gate；而 Bob 也是隨機對平均一半的量子位元操作 Hadamard Gate，另一半不操作 Hadamard Gate。因此，Alice 和 Bob 平均只會有一半的量子位元是採用相同控制訊號，所以只會有一半的量子位元是可以作為共享秘密。以圖 10 為例，假設 Alice 分別對量子位元 q[0]和 q[2]操作 NOT Gate，讓量子位元|q[3]q[2]q[1]q[0]⟩初始值為|0101⟩。Alice 分別對量子位元 q[3]和 q[2]選擇不採用 Hadamard Gate，並且分別對量子位元 q[1]和 q[0]選擇採用 Hadamard Gate，然後傳送量子位元|01UU⟩，U 表示均勻量子疊加態。之後，Bob 分別對量子位元 q[3]和 q[1]選擇不採用 Hadamard Gate，並且分別對量子位元 q[2]和 q[0]選擇採用 Hadamard Gate，所以 Bob 量測後可以得到|0UU1⟩。由於 Alice 和 Bob 在量子位元 q[3]和 q[0]都採用相同的控制訊號，所以 Alice 和 Bob 在量子位元 q[3]和 q[0]可以得到一致的量測結果。需要注意的是，在控制訊號也可以改為本研究提出的採用或不採用 SX Gate 和採用或不採用 Inversed SX Gate 兩種可選的控制訊號來達到相同效果。

## IV. E91 協定

第 IV.A 節介紹 E91 協定的流程及步驟，第 IV.B 節介紹本研究採用的控制訊號(或稱"測量基")。然後，第 IV.C 節和第 IV.D 節分別介紹通訊雙方選擇相同或不同控制訊號的結果。最後，第 IV.E 節提供案例和討論。

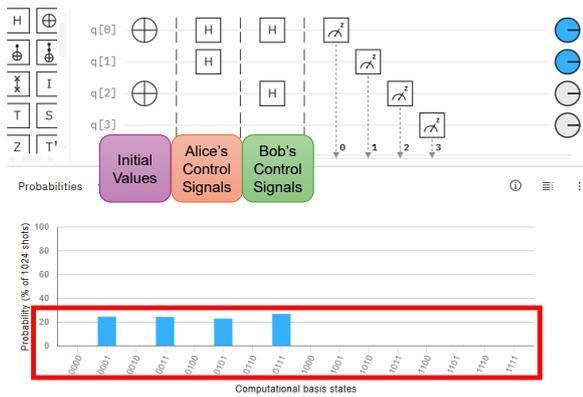

Fig. 10. A case of the BB84 protocol.

A. *步驟*

本節假設 E91 協定的傳輸雙方分別是 Alice 和 Bob。首先，先為兩兩量子位元建立貝爾態(如圖 11❶所示)，之後分別讓 Alice 和 Bob 各別持有處於貝爾態的量子位元(如圖 11❷所示)。然後，Alice 隨機選擇她的控制訊號，並且對每個量子位元進行操作，部分處於貝爾態的量子位元將會成為均勻量子疊加態(如圖 11❸所示)。Bob 也隨機選擇他的控制訊號，並且對每個量子位元進行操作(如圖 11❹所示)。最後，Alice 和 Bob 再各自量測其持有的量子位元(如圖 11❺所示)，以及傳送和比對控制訊號序列(如圖 11❻、圖 11❼、圖 11❽所示)，則後續雙方可以取得具有相同控制訊號的量子位元值作為協商後的共享秘密。詳細數學證明可以參考[14]。

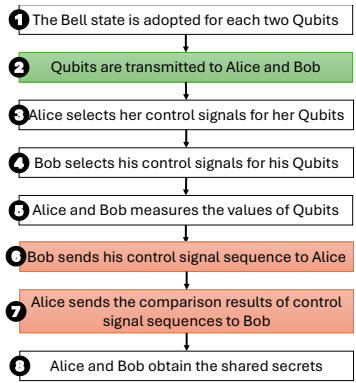

Fig. 11. The steps of the E91 protocol.

B. *控制訊號(或稱"測量基")*

本節採用兩種設計方法：(1).基於 Hadamard Gate 和(2).基於 SX Gate (本研究提出方法)。

  *1)* *基於 Hadamard Gate*：採用 Hadamard Gate 和不採用 Hadamard Gate 兩種可選的控制訊號。Alice 和 Bob 可以對各自的量子位元隨機選擇是否採用 Hadamard Gate 和不採用 Hadamard Gate。以圖 12 為例，首先，為量子位元 q[1]和 q[0]建立貝爾態(也就是|00⟩和|11⟩的機率各 50%)，並且讓 Alice 持有量子位元 q[0]，而 Bob 持有量子位元 q[1]。之後，Alice 選擇對量子位元採用 Hadamard Gate 操作(即 H)，所以量子位元 q[1]和 q[0]將會是均勻量子疊加態，也就是|00⟩、|01⟩、|10⟩、|11⟩的機率各 25%。之後，Bob 也選擇對量子位元採用 Hadamard Gate 操作(即 H)，所以量子位元還原為|00⟩和|11⟩的機率各 50%。因此，之後量測時根據量子糾纏態可以確定 Alice 和 Bob 量測的結果會是一致的。由此可知，雙方對該量子位元採用相同的控制訊號，所以該量子位元的量測結果可用。

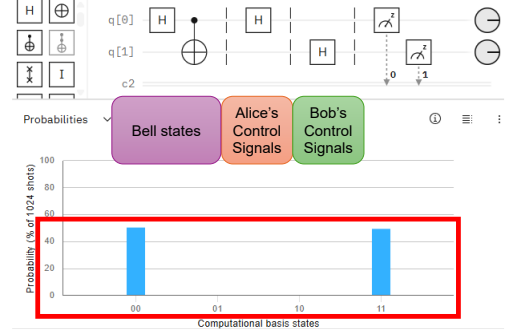

Fig. 12. The E91 protocol based on Hadamard gates.

  *2)* *基於 SX Gate*：對 Alice 有採用 SX Gate 和不採用 SX Gate 兩種可選的控制訊號；對 Bob 有採用 Inversed SX Gate 和不採用 Inversed SX Gate 兩種可選的控制訊號。Alice 和 Bob 可以隨機選擇各自的控制訊號對各自的量子位元進行操作。以圖 13 為例，首先，為量子位元 q[1]和 q[0]建立貝爾態(也就是|00⟩和|11⟩的機率各 50%)，並且讓 Alice 持有量子位元 q[0]，而 Bob 持有量子位元 q[1]。之後，Alice 選擇對量子位元採用 SX Gate 操作(即 √X)，所以量子位元 q[1]和 q[0]將會是均勻量子疊加態，也就是|00⟩、|01⟩、|10⟩、|11⟩的機率各 25%。之後，Bob 也選擇對量子位元採用 Inversed SX Gate 操作(即 √X†)，所以量子位元還原為|00⟩和|11⟩的機率各 50%。因此，之後量測時根據量子糾纏態可以確定 Alice 和 Bob 量測的結果會是一致的。由此可知，雙方對該量子位元採用相同的控制訊號，所以該量子位元的量測結果可用。

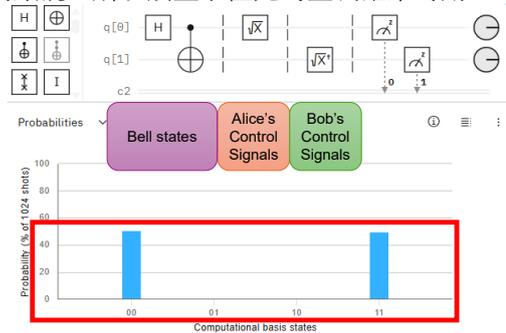

Fig. 13. The E91 protocol based on SX gates and inversed SX gates.

C. *通訊雙方選擇相同控制訊號*

本節介紹當 Alice 和 Bob 通訊雙方選擇相同控制訊號後的量測結果，如圖 14 所示。首先，建立量子位元 q[0]和 q[2]的貝爾態，以及建立量子位元 q[1]和 q[3]的貝爾態。由 Alice 持有量子位元 q[0]和 q[1]，而 Bob 持有量子位元 q[2]和 q[3]。Alice 分別對量子位元 q[0]和 q[1]選擇採用 Hadamard Gate，然後量子位元將變成|UUUU⟩，U 表示均勻量子疊加態。之後，Bob 分別對量子位元 q[2]和 q[3]選擇採用 Hadamard Gate，量子位元將變成|0000⟩、|0101⟩、|1010⟩、|1111⟩的機率各 25%。可以觀察到，在量子糾纏態的基礎上，q[0]和 q[2]會有相同的量測結果，並且 q[1]和 q[3] 會有相同的量測結果。由此可知，Alice 和 Bob 在其各自處於貝爾態的量子位元都採用相同的控制訊號時，Alice 和 Bob 在量測各自的量子位元可以得到一致的量測結果。

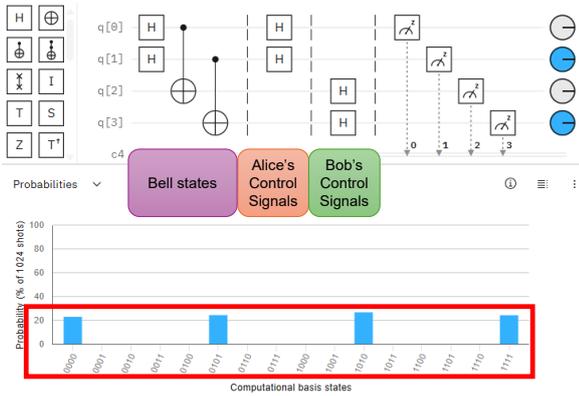

Fig. 14. Selection of the same control signals by both communication parties in the E91 protocol.

### D. 通訊雙方選擇不同控制訊號

本節介紹當 Alice 和 Bob 通訊雙方選擇不同控制訊號後的量測結果,如圖 15 所示。首先,建立量子位元 q[0] 和 q[2]的貝爾態,以及建立量子位元 q[1]和 q[3]的貝爾態。由 Alice 持有量子位元 q[0]和 q[1],而 Bob 持有量子位元 q[2]和 q[3]。Alice 分別對量子位元 q[0]選擇採用 Hadamard Gate,q[1]選擇不採用 Hadamard Gate。之後,Bob 分別對量子位元 q[2]選擇不採用 Hadamard Gate,q[3]選擇採用 Hadamard Gate,量子位元將變成|UUUU〉。由此可知,Alice 和 Bob 在其各自處於貝爾態的量子位元都採用不同的控制訊號時,Alice 和 Bob 在量測各自的量子位元無法保障與對方得到一致的量測結果。

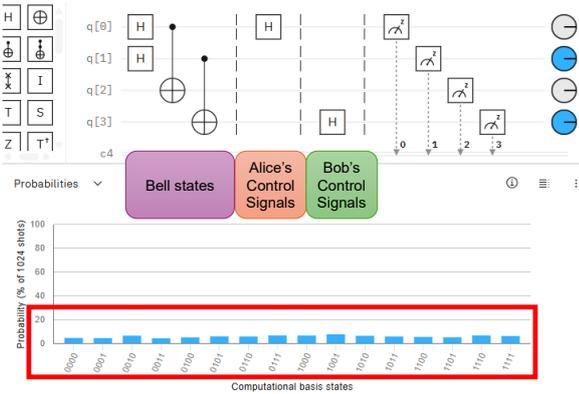

Fig. 15. Selection of the different control signals by both communication parties in the E91 protocol.

### E. 案例與討論

在實際部署 E91 協定時,控制訊號的部分,如果都選擇操作 Hadamard Gate 雖然可以讓全部量子位元在傳輸過程中都是均勻量子疊加態,但攻擊者只需對全部量子位元都 Hadamard Gate 即可還原。因此,為了提升傳輸安全和增加傳輸值的亂度,Alice 隨機對其持有的平均一半的量子位元操作 Hadamard Gate,另一半不操作 Hadamard Gate;而 Bob 也是隨機對其持有的平均一半的量子位元操作 Hadamard Gate,另一半不操作 Hadamard Gate。因此,Alice 和 Bob 平均只會各自持有的一半的量子位元是採用相同控制訊號,所以只會有各自持有的一半的量子位元是可以作為共享秘密。以圖 16 為例,首先,建立量子位元 q[0]和 q[2]的貝爾態,以及建立量子位元 q[1]和q[3]的貝爾態。由Alice持有量子位元 q[0]和q[1],而 Bob 持有量子位元 q[2]和 q[3]。假設 Alice 分別對量子位元 q[0]和 q[1]選擇採用 Hadamard Gate。之後,Bob 分別對量子位元 q[2]選擇採用 Hadamard Gate,q[3]選擇不採用 Hadamard Gate,所以量子位元將變成|U0U0〉或 |U1U1〉,U 表示均勻量子疊加態。由於量子位元 q[0]和 q[2]處於量子糾纏態,並且 Alice 和 Bob 在該量子位元採用相同控制訊號,所以 q[0]和 q[2]將得到一致的量測結果。需要注意的是,在控制訊號也可以改為本研究提出的採用或不採用 SX Gate 和採用或不採用 Inversed SX Gate 兩種可選的控制訊號來達到相同效果。

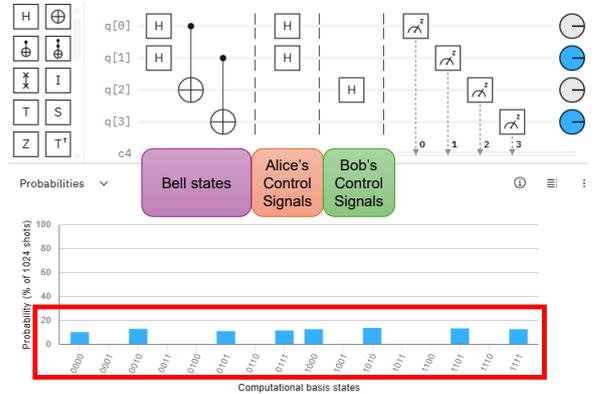

Fig. 16. A case of the E91 protocol.

## V. 實證結果與討論

本節講述本研究的實驗環境和驗證結果討論。其中,第 V.A 節介紹實驗環境,第 V.B 節~第 V.D 節分別討論熵驗證、獨立且同分布驗證、錯誤率驗證的結果。

### A. 實驗環境

為驗證量子金鑰分配,本研究採用 Python 3.11.14 和 IBM Qiskit 2.2.3 進行實作,並實際於 IBM Quantum Platform [15]以真實量子電腦執行和實證。其中,由於 IBM Quantum Platform 目前提供的量子處理單元的量子位元數上限分別為 133 個或 156 個(如圖 17 所示),所以本研究設計的量子電路以 128 個量子位元數為主。

| QPU name | Instance | Qubits | Status | Pending jobs |
|---|---|---|---|---|
| ibm_boston | | 156 | ● Online | 0 |
| ibm_kingston | | 156 | ● Online | 4 |
| ibm_pittsburgh | | 156 | ● Online | 607 |
| ibm_fez | open-instance | 156 | ● Online | 55 |
| ibm_marrakesh | open-instance | 156 | ● Paused - maintenance | 19,992 |
| ibm_torino | open-instance | 133 | ● Online | 4 |

Fig. 17. The computation resources of IBM Quantum Platform on Dec. 9, 2025 on the webpage (i.e. https://quantum.cloud.ibm.com/computers).

### B. 熵驗證

為驗證量子金鑰分配協商後的共享秘密的隨機性,本研究採用 NIST SP 800-90B [16]的熵計算方法並且進行二項式分布檢定,得到多組共享秘密計算其熵值和 p-value。其中,熵值和 p-value 越大表示效能越好,並且 NIST SP 800-90B 定義 p-value 門檻為 0.000005。由實驗結果顯示,不論是BB84協定或E91協定所得到的共享秘密的p-value 皆大於 0.000005,表示量測到的 0 或 1 的機率接近各 50%,可以通過熵驗證,如表 I 所示。

TABLE I. ENTROPY VALIDATION

| QKD Protocol | Entropy | p-value |
|---|---|---|
| BB84 Protocol Based on Hadamard Gate | 0.982 | 0.260 |
| BB84 Protocol Based on SX Gate (The proposed method) | 0.988 | 0.381 |
| E91 Protocol Based on Hadamard Gate | 0.857 | 0.020 |
| E91 Protocol Based on SX Gate (The proposed method) | 0.989 | 0.597 |

*C. 獨立且同分布驗證*

NIST SP 800-90B [16]規範獨立且同分布(Independent and Identically Distributed, IID) 驗證包含獨立性(Independent, IND)檢定、適合度(Goodness of Fit, GOF)檢定，以及最長重覆子字串(Longest Repeated Substring, LRS)檢定，並且 p-value 門檻為 0.001。表 II 為獨立且同分布驗證，實驗結果顯示不論是 BB84 協定或 E91 協定所得到的共享秘密在各個驗證指標的 p-value 皆大於 0.001，可以通過獨立且同分布驗證。

TABLE II. INDEPENDENT AND IDENTICALLY DISTRIBUTED VALIDATION

| QKD Protocol | IND Test p-value | GOF Test p-value | LRS Test p-value |
|---|---|---|---|
| BB84 Protocol Based on Hadamard Gate | 0.291 | 0.386 | 0.998 |
| BB84 Protocol Based on SX Gate (The proposed method) | 0.333 | 0.002 | 1.000 |
| E91 Protocol Based on Hadamard Gate | 0.139 | 0.011 | 0.996 |
| E91 Protocol Based on SX Gate (The proposed method) | 0.333 | 0.157 | 1.000 |

*D. 錯誤率驗證*

為驗證在真實量子電腦環境執行可能造的錯誤率，本研究實證結果如表 III 所示。其中，可以觀察到 BB84 協定僅用到量子疊加態，每個量子位元可獨立運作，所以目前錯誤率已經接近 0。然而，E91 協定需建構在量子糾纏態的基礎，但目前在量測量子糾纏態時的誤差仍大約有一成錯誤率。

TABLE III. ERROR RATE VALIDATION

| QKD Protocol | Error Rate |
|---|---|
| BB84 Protocol Based on Hadamard Gate | 0.000 |
| BB84 Protocol Based on SX Gate (The proposed method) | 0.000 |
| E91 Protocol Based on Hadamard Gate | 0.125 |
| E91 Protocol Based on SX Gate (The proposed method) | 0.094 |

## VI. 結論與未來研究

本研究在真實量子電腦實作 BB84 協定和 E91 協定等量子金鑰分配協定，並且提供基於 SX Gate 的量子金鑰分配協定。由實證結果顯示所提方法通過從熵、獨立且同分布、以及錯誤率等驗證，證明其可行性。然而，目前量子糾纏態的量測錯誤率仍未接近 0，未來可朝量子糾纏態的量測技術改進。